\documentclass[%
 reprint,
superscriptaddress,
 amsmath,amssymb,
 aps,
floatfix,
prd
]{revtex4-2}
\usepackage{booktabs}  
\usepackage{graphicx}
\usepackage{dcolumn}
\usepackage{bm}
\usepackage{float}
\usepackage[usenames,dvipsnames]{color}
\usepackage{caption}
\usepackage{subcaption}

\begin{document}
\newcommand{\bea}{\begin{eqnarray}}
\newcommand{\eea}{\end{eqnarray}}
\newcommand{\beq}{\begin{equation}}
\newcommand{\eeq}{\end{equation}}
\def\leq{\raise 0.4ex\hbox{$<$}\kern -0.8em\lower 0.62ex\hbox{$-$}}
\def\geq{\raise 0.4ex\hbox{$>$}\kern -0.7em\lower 0.62ex\hbox{$-$}}
\def\lsim{\raise 0.4ex\hbox{$<$}\kern -0.75em\lower 0.65ex\hbox{$\sim$}}
\def\gsim{\raise 0.4ex\hbox{$>$}\kern -0.75em\lower 0.65ex\hbox{$\sim$}}
\def\pm{\,\raise 0.4ex\hbox{$+$}\kern -0.75em\lower 0.65ex\hbox{$-$}\,}

\preprint{APS/123-QED}

\title{Modeling Relative Peak Times of Gravitational Wave Harmonics}

\author{Anuj Kankani}
\email{anuj.kankani@mail.wvu.edu}
\affiliation{Department of Physics and Astronomy, West Virginia University, Morgantown, WV 26506, USA
}
\affiliation{Center for Gravitational Waves and Cosmology, West Virginia University, Morgantown, WV 26505, USA
}
\author{Sean T. McWilliams}
\affiliation{Department of Physics and Astronomy, West Virginia University, Morgantown, WV 26506, USA
}
\affiliation{Center for Gravitational Waves and Cosmology, West Virginia University, Morgantown, WV 26505, USA
}

\date{\today}

\begin{abstract}
Accurate modeling of gravitational waves from binary black hole mergers is essential for extracting their rich physics. A key detail for understanding the physics of mergers is predicting the precise time when the amplitude of the gravitational wave strain peaks, which can differ significantly among the different harmonic modes. We propose two semi-analytical methods to predict these differences using the same three inputs from Numerical Relativity (NR): the remnant mass and spin and the instantaneous frequency of each mode at its peak amplitude. The first method uses the frequency evolution predicted by the Backwards-One-Body model, while the second models the motion of an equatorial timelike geodesic in the remnant black hole spacetime. We compare our models to the SXS waveform catalog for quasi-circular, non-precessing systems and find excellent agreement for $l = |m|$ modes up to $l=8$, with mean and median differences from NR below 1$M$ in nearly all cases across the parameter space. We compare our results to the differences predicted by leading Effective-One-Body and NR surrogate waveform models and find that in cases corresponding to the largest timing differences, our models can provide significant increases in accuracy.
\end{abstract}

\maketitle


\section{Introduction}
The era of gravitational-wave (GW) astronomy has rapidly transformed our understanding of the universe, from revealing the formation and evolution of black holes \cite{evol_cite1,evol_cite2} to enabling strong-field tests of General Relativity (GR) \cite{strong_field_GR_test}. Critical to these advances is the ability to accurately model the GWs from merging compact objects. The increased sensitivity of future detectors such as LISA \cite{LISA,LISA_cite2}, Cosmic Explorer \cite{CE_cite1,CE_cite2}, and other advanced ground-based detectors \cite{LIGO_cite1,LIGO_cite2} will require waveform models that incorporate harmonics beyond the dominant $(l, |m|) = (2, 2)$ mode \cite{higher_mode_cite1,higher_mode_cite2}.

The inclusion of higher-order modes improves the determination of intrinsic source parameters \cite{higher_mode_cite2,higher_mode_degen1} and source luminosity distance and inclination \cite{higher_mode_degen2,higher_mode_distance1}, leading to better constraints on black hole properties and their formation channels. Additionally, accurate modeling of higher modes can provide more stringent tests of GR \cite{higher_mode_grtest1,higher_mode_grtest2,higher_mode_grtest3}.

Numerical relativity (NR) simulations currently provide the most accurate waveforms, but are too computationally expensive for direct use in parameter estimation. Therefore, various approaches combine NR's accuracy with the efficiency of analytical or data-driven methods. NR surrogate models \cite{surr_cite1,surr_cite2} directly interpolate NR waveforms. Models based on the Effective-One-Body (EOB) formalism \cite{EOB_cite1,EOB_cite2,SEOBNR_cite1,SEOBNR_cite2,TEOB_cite1,TEOB_cite2} combine Post-Newtonian (PN) theory \cite{PN_cite1,PN_cite2,PN_cite3} with NR information to create accurate and efficient waveform approximants. Alternative approaches, such as the Backwards-One-Body (BOB) model \cite{BOB}, combine EOB-based inspirals with an analytical merger-ringdown, greatly reducing reliance on extensive phenomenological calibration \cite{Mahesh:2025}.

In this paper, we investigate the problem of predicting the time of peak strain amplitude for various GW modes. Although all modes peak within a small time window compared to the full IMR waveform duration, their precise peak times differ non-negligibly over the timescale of the merger. Capturing these differences is crucial for accurately modeling subdominant modes, especially since these peaks occur in the strong-field merger phase where PN expansions are insufficient, forcing a significant reliance on NR. We introduce a semi-analytical approach that leverages existing NR inputs more efficiently. We demonstrate that key NR information already used in IMR models, such as \cite{Mahesh:2025}, can be combined with analytical methods to predict other strong-field observables, in particular the relative times at which different modes reach their peak amplitudes. These approaches can reduce the extensive use of NR in modeling the merger, and ensure NR information can be leveraged where it is most needed. We introduce two semi-analytical methods to predict relative peak times without requiring extensive calibration against NR catalogs. We verify our methods by comparing our predictions to results obtained directly from NR simulations.

\section{Methods}
We consider two methods to determine the peak time differences of various $l=|m|$ modes for quasi-circular, non-precessing systems. Both methods require the same NR inputs: the remnant black hole mass $M_f$, the remnant spin $\chi_f$, and the instantaneous waveform frequency of each $(l,m)$ mode at its peak strain amplitude. While various analytical estimates for remnant properties exist \cite{final_mass,final_mass_eob,BOB}, for simplicity we use values from NR simulations.

\subsection{The Backwards-One-Body Model}
Our first method uses the closed-form frequency evolution predicted by the Backwards-One-Body (BOB) model \cite{BOB}:
\begin{equation}
    \Omega(t) = \bigg(\Omega_0^4 + k\bigg[\tanh\bigg(\frac{t-t_p}{\tau}\bigg) - \tanh\bigg(\frac{t_0-t_p}{\tau}\bigg)\bigg]\bigg)^{\frac{1}{4}},
\end{equation}
where $k$ is defined as 
\begin{equation}
    k = \frac{\Omega_\text{QNM}^4 - \Omega_0^4}{1-\tanh(\frac{t_0-t_p}{\tau})},
\end{equation}
and $\Omega_\text{QNM} = \omega_\text{QNM}/m$. Here, $\omega_\text{QNM}$ is the real part of the dominant $(l, m, n) = (2, 2, 0)$ \footnote{$n$ denotes the overtone number; we will assume that only the least damped fundamental overtone $n=0$ contributes, and denote modes as $(l, m)$ hereafter.} quasinormal mode (QNM) frequency, and the damping time $\tau$ is the inverse of its imaginary part. The waveform frequency is then approximated by $\omega \approx m\Omega$.

We set the peak of the $(2, 2)$ strain to occur at $t_0 = 0$. The peak of $\psi_4$ for the $(2, 2)$ mode, denoted $t_p$, is then predicted by BOB to occur at:
\begin{equation}
    t_p = -2\tau \ln\bigg(\frac{\Omega_0}{\Omega_\text{QNM}}\bigg),
\end{equation}
where we set $\Omega_0 = \omega_{22p}/2$, with $\omega_{22p}$ being the strain frequency at the peak amplitude of the $(2, 2)$ mode. Although this evolution is constructed for the Weyl scalar $\psi_4$, we assume the same underlying frequency evolution applies to the strain, $h$. Therefore, given the peak strain frequency of any other $(l, |m|=l)$ mode from NR, we can use Eq.~(1) to determine the time at which that frequency is reached, thereby predicting the mode's peak time. The BOB framework can be used to construct various closed form expressions for the frequency evolution, based on different assumptions and initial conditions. While all the different expressions show excellent amplitude and frequency agreement with NR, the very small timing differences necessitate constructing BOB for maximal agreement near the merger. We tested different methods for constructing BOB and found this formulation provided the best agreement with NR.

\subsection{The Geodesic Model}
Our second model provides an alternative prediction without relying on a frequency evolution model. We associate the peak emission of each GW mode with a test particle crossing a characteristic radius, $r_{lm}$, in the remnant black hole spacetime. This is inspired by one-body models where strong-field events, such as crossing the light ring, correspond to GW features, and by the distinct effective potentials experience by different modes of scalar radiation \cite{bardeen}. We restrict our analysis to equatorial geodesics ($Q=0$), consistent with our focus on non-precessing systems and $l = |m|$ modes. We correlate the geodesic motion with the peak waveform amplitude as follows:

1. Convert the NR strain frequency $\omega_{lm}$ at the peak of each mode to an orbital frequency: 
\beq
\Omega = \omega/m.
\eeq

2. Calculate the corresponding circular orbit radius in the remnant spacetime \cite{bardeen}:
\beq
\Omega = \frac{\sqrt{M_f}}{r^{\frac{3}{2}} + a\sqrt{M_f}},
\label{eq:radius}
\eeq
where we assume prograde orbits and $a = |\chi_f|M_f$. Solving for $r$ yields the characteristic radius $r_{lm}$ associated with the peak of each $(l, m)$ mode.\\

3. Select the specific energy $E/\mu$ and angular momentum $L/(\mu M_f)$ that define the geodesic's trajectory. A geodesic plunging from the Innermost Stable Circular Orbit (ISCO) \cite{bardeen} of the remnant black hole is a physically motivated choice. However, the characteristic radii $r_{lm}$ can be larger than the ISCO radius, $r_\text{ISCO}$, meaning a geodesic plunging from $r_\text{ISCO}$ would have already passed $r_{lm}$. We find that rescaling the ISCO energy and angular momentum, $E_{\text{ISCO}}/\mu$ and $L_{\text{ISCO}}/\mu$, by the factor $M_i/M_f$, where $M_i$ is the binary's initial mass, provides a trajectory that correlates well with NR peak timings. This empirical rescaling may account for energy lost during the inspiral. For binaries with mass ratio $q \,\leq\, 1.5$, we find it is better to use the remnant mass estimated at the ISCO crossing from \cite{BOB} as the background spacetime mass. For simplicity, in this case, we retain the final remnant spin and only adjust the mass.\\

4. Calculate the coordinate time for a timelike geodesic to travel between two characteristic radii, $r_{lm,1}$ and $r_{lm,2}$, by solving the geodesic equations of motion for the remnant black hole \cite{bardeen}. This time gives the predicted difference between the mode peak times.\\

\begin{figure}[h!]
    \centering
    \includegraphics[width=0.45\textwidth]{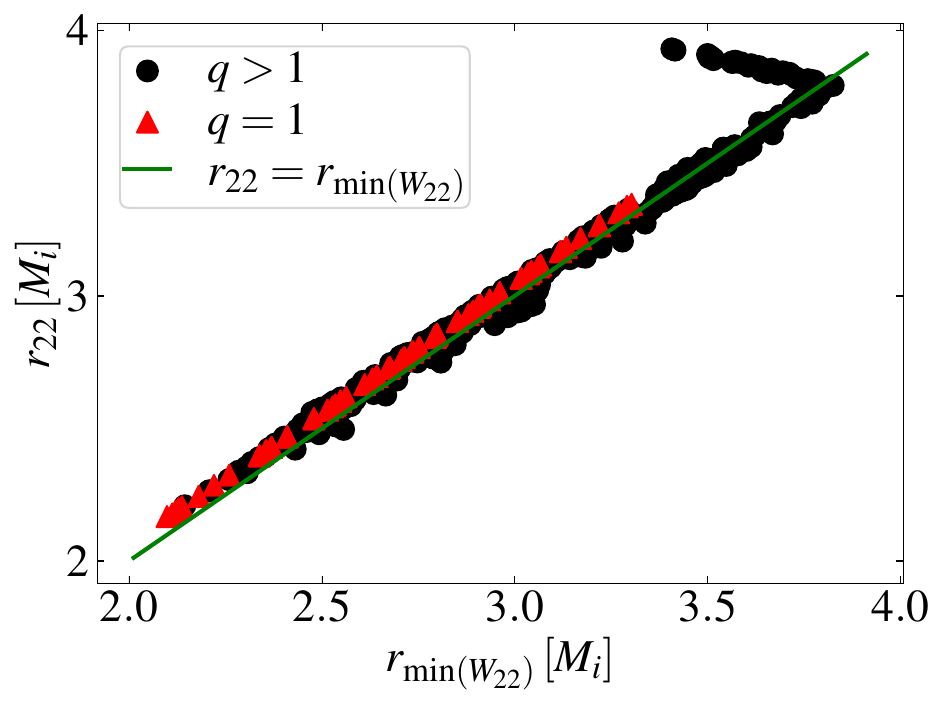}
    \caption{Characteristic radii $r_{22}$ calculated using Eq. (\ref{eq:radius}) as a function of the radius at which the minimum of the scalar effective potential $W$ occurs for the $(2,2)$ mode. For reference, an exact agreement between the two quantities is shown by a green line. Equal mass configurations are shown with filled red triangles, all other configurations are shown with filled black circles. The kink seen in the upper right may be due to poor NR frequency data, resulting in deviations from the linear trend.}
    \label{fig:scalar_22}
    \includegraphics[width=0.45\textwidth]{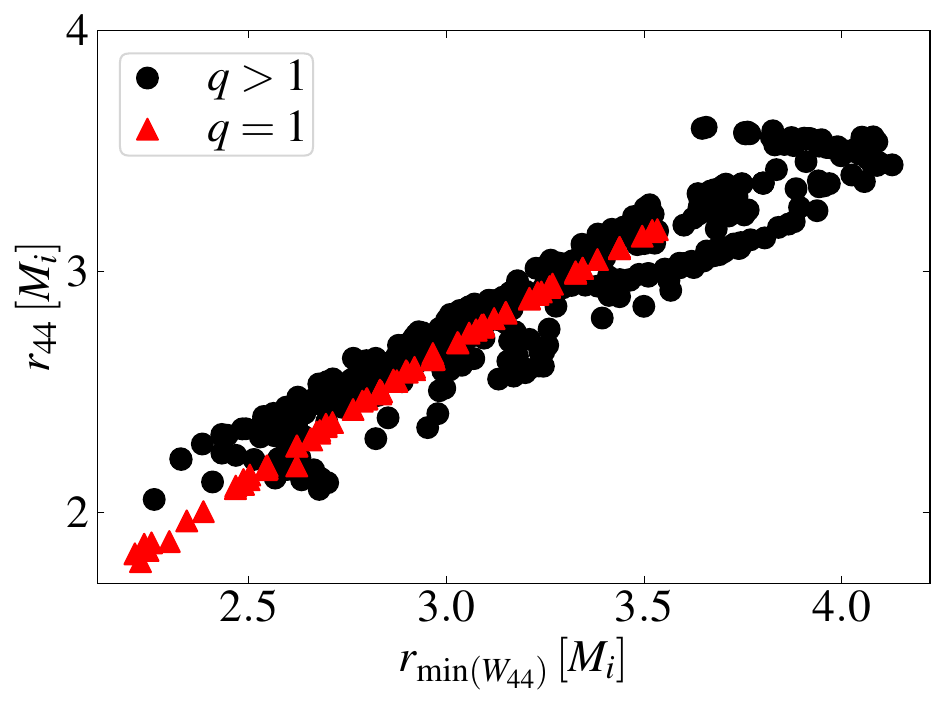}
    \caption{Characteristic radii $r_{44}$ calculated using Eq. (\ref{eq:radius}) as a function of the radius at which the minimum of the scalar effective potential $W$ occurs for the $(4,4)$ mode. Equal mass configurations are shown with filled red triangles, all other configurations are shown with filled black circles.}
    \label{fig:scalar_44}
\end{figure}

We note that, in an effort to eliminate our dependence on NR for the frequencies of each mode at their peak amplitudes, we also investigated the connection between our characteristic radii and the effective potentials that govern scalar and tensor wave propagation in a Kerr background. We first compared the characteristic radii used in this work against the minimum of the mode-dependent effective potential for scalar radiation \cite{bardeen}. In Figure~\ref{fig:scalar_22}, we show that for the $(2, 2)$ mode these values are almost identical. For higher modes, the characteristic radii for equal mass configurations show a linear relationship with the potential minimum, but with a non-zero intercept, and with some nonnegligible scatter around this relationship if we include configurations with higher mass ratios; in Figure~\ref{fig:scalar_44} we show the $(4, 4)$ mode as an example. We also compared our characteristic radii with the minima of the more complicated tensor potentials \cite{chandra} and found similar results. There is an obvious ambiguity in this comparison since any estimate of a radius will be gauge dependent. Nonetheless, the simple relationships between the characteristic radii and the minima of these effective potentials suggest that there may be a rigorous analytical approach to obtain these radii directly, and it may therefore be possible to remove the direct dependence on NR frequency, but we leave further study of this possibility for future work.

\subsection{Error Estimation}
\label{subsec:Error}
To estimate the NR error, we use the highest and second-highest resolutions available. We define $\delta t_{\text{NR}}$ as the absolute difference in the peak timings calculated from these two resolutions. To ensure data quality, we only include SXS simulations where $\delta t_{\text{NR}} < 1M_i$. Since our models use the NR frequency at peak strain as input, we propagate this uncertainty by defining a model error, $\delta t_\text{model}$, as the difference in our prediction when using frequencies from the two highest resolutions. We only consider cases where $\delta t_\text{model} < 1M_i$. This filtering removes low-quality simulations from our comparison while retaining enough waveforms to sample the parameter space.  Additionally, five individual data points were not captured by the automated filtering and were manually removed after visual inspection due to their low waveform quality. Table~\ref{tab:bad_sims_1} in Appendix A lists the specific SXS ID numbers and modes that were removed from the analysis.

\section{Results}
We now compare our BOB-based and geodesic models for the relative peak times of each mode against results from the SXS waveform catalog \cite{sxs_cat1,sxs_cat2,sxs_cat3}. All comparisons are for quasi-circular, non-precessing systems. We apply the filtering criteria from Section~\ref{subsec:Error} to the SXS data to minimize the impact of NR truncation error.

\subsection{The Equal Mass Case}
\begin{figure*}[h!tpb]
\centering
\includegraphics[width=0.32\textwidth]{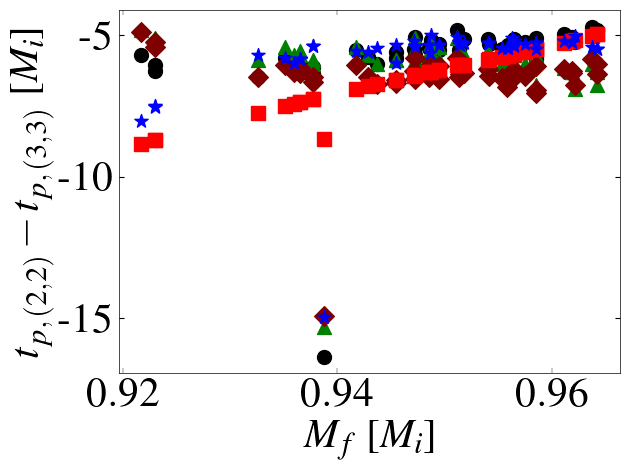}
\includegraphics[width=0.32\textwidth]{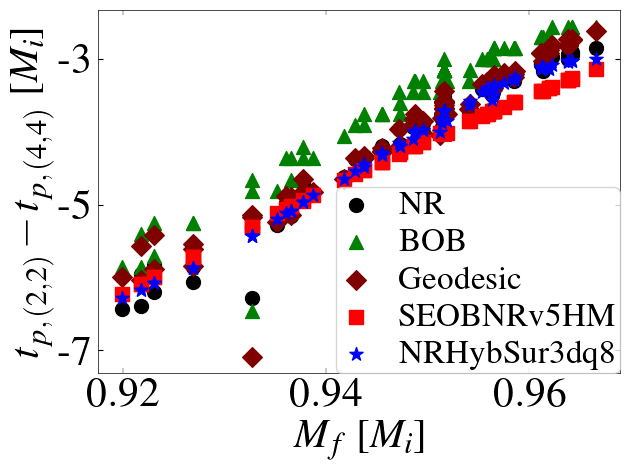}
\includegraphics[width=0.32\textwidth]{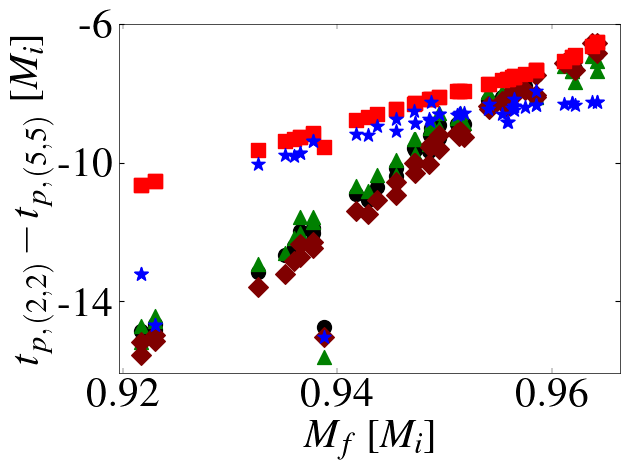}
\includegraphics[width=0.32\textwidth]{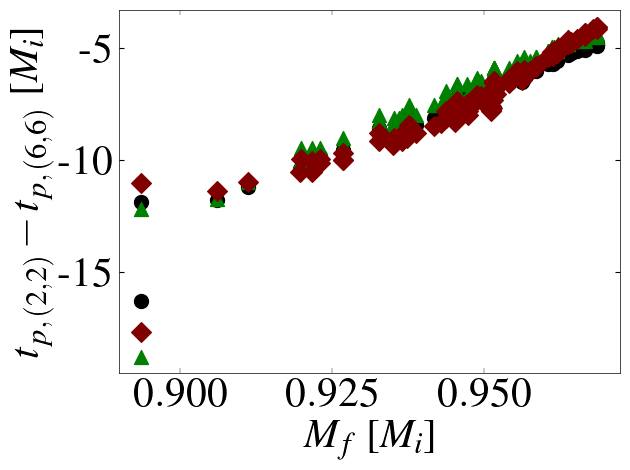}
\includegraphics[width=0.32\textwidth]{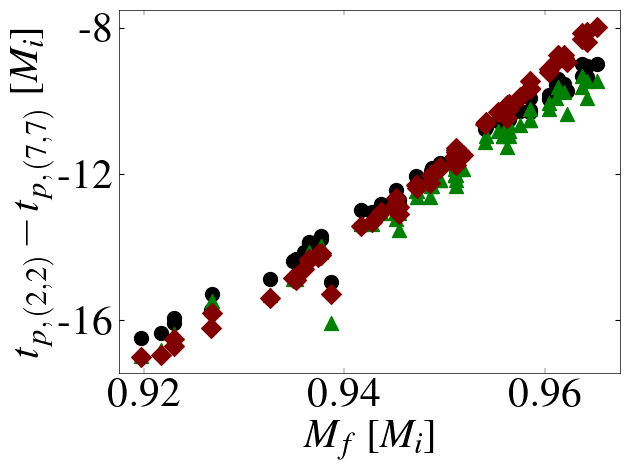}
\includegraphics[width=0.32\textwidth]{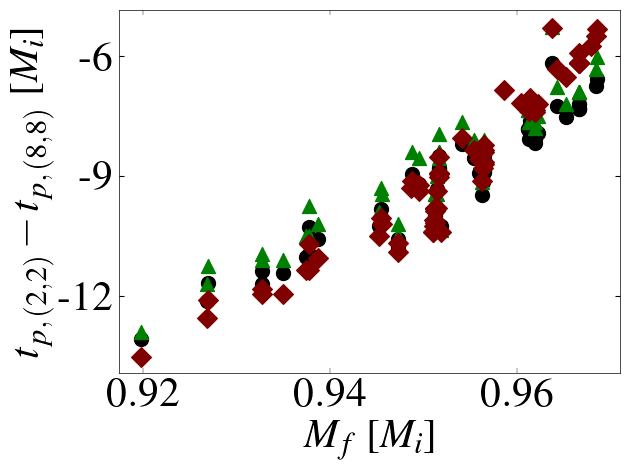}
\caption{Comparison of predicted time differences between the peak amplitude of the $(2, 2)$ and corresponding $(l, m)$ modes for equal-mass binaries as a function of the remnant mass. Results from our Geodesic model (brown diamonds), BOB model (green triangles), NRHybSur3dq8 (blue stars), and SEOBNRv5HM (red squares) are compared against SXS NR data (black circles). For $l\,\leq\,5$, we selected systems with initial dimensionless spin $|\chi_i| \,\leq \,0.8$ for consistency with NRHybSur3dq8's training limits. SEOBNRv5HM and NRHybSur3dq8 do not model $l>5$ modes so comparisons to NR are not possible for $l>5$ modes. The error on the SXS, Geodesic, and BOB data points, calculated through comparisons using lower resolution data, are not shown but are all under $1M_i$.}
\label{fig:q1_cases}
\end{figure*}

Equal-mass ($q=1$) binaries often produce the largest peak time differences across modes and serve as a stringent test case. Figure~\ref{fig:q1_cases} compares the difference in the time between the peak amplitude of the $(2, 2)$ mode and each $(l, m=l)$ mode up to $l=8$ computed by various models and NR. Due to our data quality filters, higher-$l$ modes may have fewer data points.

Table \ref{table:q1_table} quantifies the accuracy by listing the average absolute difference between model predictions and NR results for each $(l, m)$ mode. The ``NR'' column gives the average timing difference, providing a scale for the typical values. Both our models have mean and median differences $\lessapprox1M_i$ for all modes and under $0.55 M_i$ for all $l\,\geq\,4$ modes. Both the surrogate and EOB waveform models agree well for the $(4, 4)$ mode but show significant disagreement for the $(5, 5)$ modes. For the $(5, 5)$ mode, both IMR models have average differences greater than $1 M_i$ and, as seen in Figure \ref{fig:q1_cases}, fail to capture the trend or magnitude of the timing differences as accurately as our BOB and geodesic models. Both our models maintain good agreement with NR for higher-$l$ modes.

\begin{table*}[h!tpb]
\centering
\begin{tabular}{|@{\hskip 0.5cm}c@{\hskip 0.5cm}||@{\hskip 0.5cm}c@{\hskip 0.5cm}||@{\hskip 0.5cm}c@{\hskip 0.5cm}c@{\hskip 0.5cm}c@{\hskip 0.5cm}c@{\hskip 0.5cm}|}
\hline
\textbf{$(l,m)$} & \textbf{NR} $[M_i]$& \textbf{$\Delta$ Geodesic $[M_i]$} & \textbf{$\Delta$ BOB $[M_i]$} & \textbf{$\Delta$ NRHybSur3dq8 $[M_i]$} & \textbf{$\Delta$ SEOBNRv5HM $[M_i]$} \\
\hline
$(3, 3)$ & 5.44 & 1.003 & 0.52 & 0.266 & 0.698 \\
$(4, 4)$ & 3.75 & 0.046 & 0.471 & 0.066 & 0.275 \\
$(5, 5)$ & 8.95 & 0.286 & 0.232 & 0.863  & 1.063  \\
$(6, 6)$ & 6.79 & 0.222 & 0.539 & -     & -     \\
$(7, 7)$ & 11.7 & 0.36 & 0.36 & -     & -     \\
$(8, 8)$ & 9.07 & 0.346 & 0.417 & -     & -     \\
\hline
\end{tabular}
\caption{Median of the absolute value of the difference between the peak amplitude times of the $(2, 2)$ and other $(l, m)$ modes for equal-mass binaries in the SXS catalog (NR), and the average absolute error in this time difference for the four model predictions under consideration. For $l\,\leq\,5$, only systems with initial dimensionless spin $|\chi_i|\,\leq\,0.8$ per black hole were selected for consistency with NRHybSur3dq8's calibration limits. SEOBNRv5HM and NRHybSur3dq8 do not model $l>5$ modes so comparisons to NR are not possible for $l>5$ modes.}
\label{table:q1_table}
\end{table*}

\begin{figure*}[h!]
\centering
\includegraphics[]{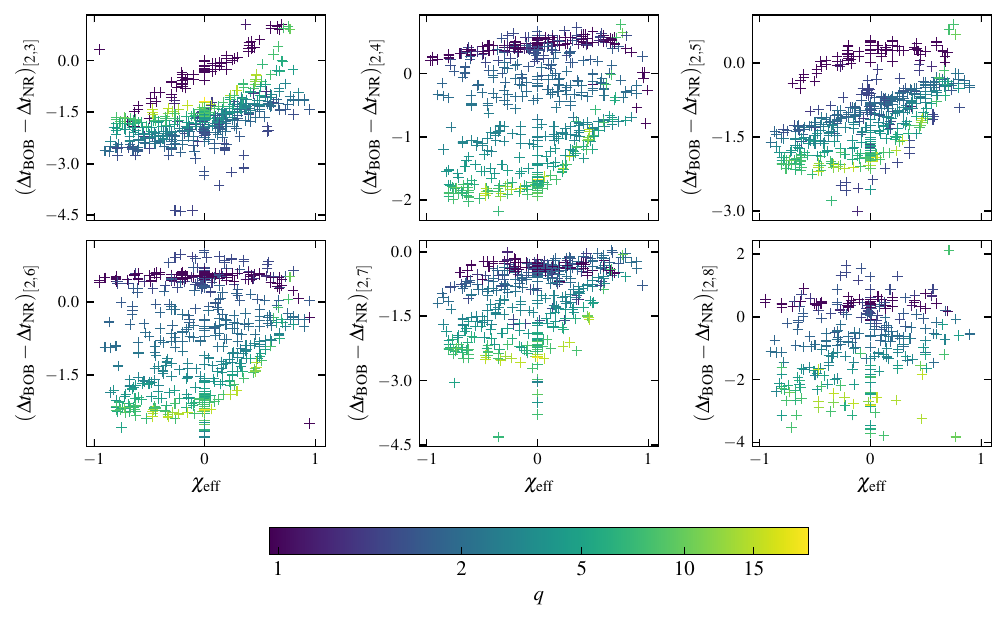}
\caption{Difference between NR and the BOB model for the time between the peak amplitude of the $(2, 2)$ and corresponding $(l, m)$ modes across all available configurations as a function of the effective spin $\chi_{\mathrm{eff}}$. The colorbar represents the initial mass ratio $q$ of each configuration.The data is in units of $M_i$ and [$l,l'$] indicates that the difference is being calculated between the $(l,m=l)$ and $(l',m'=l')$ modes. }
\label{figure:BOB_ALL}
\end{figure*}

\begin{figure*}[htpb]
\centering
\includegraphics[]{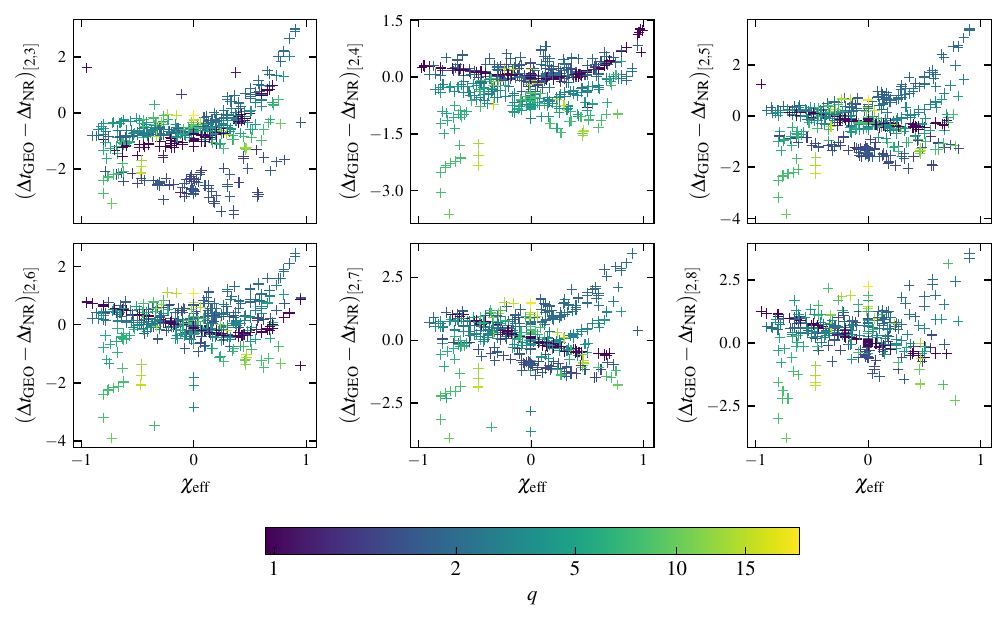}
\caption{Difference between NR and the Geodesic model for the time between the peak amplitude of the $(2, 2)$ and corresponding $(l, m)$ modes across all available configurations as a function of the effective spin $\chi_{\mathrm{eff}}$. The colorbar represents the initial mass ratio $q$ of each configuration. The data is in units of $M_i$ and [$l,l'$] indicates that the difference is being calculated between the $(l,m=l)$ and $(l',m'=l')$ modes.}
\label{figure:GEO_ALL}
\end{figure*}

\begin{figure*}[htpb]
\centering
\includegraphics[]{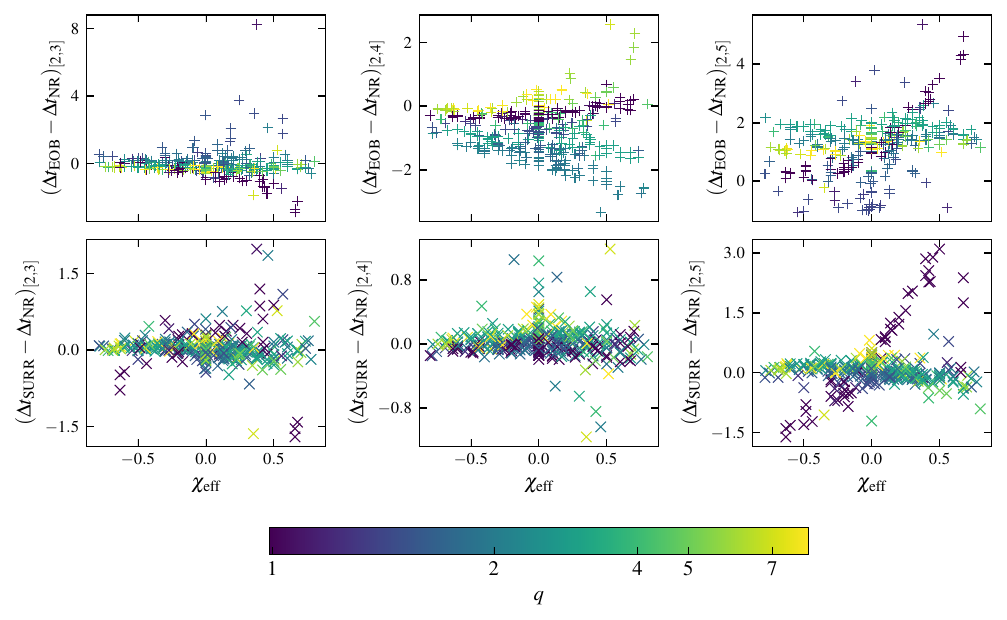}
\caption{Differences between NR and the SEOBNRv5HM (upper) and NRHybSur3dq8 (lower) IMR models for the time between the peak amplitude of the $(2, 2)$ and corresponding $(l, m)$ modes across many configurations as a function of the effective spin $\chi_{\mathrm{eff}}$. Systems are restricted to $|\chi_i|\,\leq\,0.8$ and $q\,\leq\,8$ for consistency with NRHybSur3dq8's training set. The colorbar represents the initial mass ratio $q$ of each configuration.}
\label{figure:OTHER_ALL}
\end{figure*}

\begin{table*}[h!]
\centering
\begin{tabular}{|@{\hskip 0.5cm}c@{\hskip 0.5cm}||@{\hskip 0.5cm}c@{\hskip 0.5cm}||@{\hskip 0.5cm}c@{\hskip 0.5cm}c@{\hskip 0.5cm}c@{\hskip 0.5cm}c@{\hskip 0.5cm}|}
\hline
\textbf{$(l,m)$} & \textbf{NR} $[M_i]$& \textbf{$\Delta$ Geodesic $[M_i]$} & \textbf{$\Delta$ BOB $[M_i]$} & \textbf{$\Delta$ NRHybSur3dq8 $[M_i]$} & \textbf{$\Delta$ SEOBNRv5HM $[M_i]$} \\
\hline
$(3, 3)$ & 4.51 & 0.698 & 1.689 & 0.095 & 0.276 \\
$(4, 4)$ & 2.9 & 0.229 & 0.524 & 0.08 & 0.691 \\
$(5, 5)$ & 5.05 & 0.422 & 1.01 & 0.163 & 1.408 \\
$(6, 6)$ & 5.12 & 0.31 & 0.665 & -   & -   \\
$(7, 7)$ & 6.23 & 0.615 & 0.766 & -   & -   \\
$(8, 8)$ & 6.56 & 0.589 & 0.7 & -     & -     \\
\hline
\end{tabular}
\caption{Median of the absolute value of the difference between the peak amplitude times of the $(2, 2)$ and other $(l, m)$ modes across all cases in the SXS catalog (NR), and the median absolute error in this time difference for the four model predictions under consideration. For $l\,\leq\,5$, systems are restricted to $|\chi_i|\,\leq\,0.8$ and $q\,\leq\,8$ for consistency with NRHybSur3dq8. SEOBNRv5HM and NRHybSur3dq8 do not model $l>5$ modes so comparisons to NR are not possible for $l>5$ modes.}
\label{qall_table}
\end{table*}

\subsection{All Cases}
We now extend our analysis beyond equal mass to all available mass ratios. Figure~\ref{figure:BOB_ALL} provides the differences in the timings predicted by BOB and those given by NR. We find good agreement for $l\,\geq\,4$ modes but some disagreement for the $(3, 3)$ mode, possibly due to mode-mixing effects (see e.~g.~\cite{Kelly12} and references therein).
Additionally, the BOB model tends to perform worse as the mass ratio increases. This likely stems from our dual approximation of $\omega \approx m\Omega$ and using a single frequency evolution for both $\psi_4$ and $h$.

Figure~\ref{figure:GEO_ALL} compares the geodesic model and NR data. Again, the largest disagreement is for the $(3, 3)$ mode, although this difference is smaller than for the BOB model. Nonetheless, the agreement is still excellent for all $l\,\geq\,4$ modes, and we do not see this accuracy degrade significantly with higher mass ratios. Figure~\ref{figure:OTHER_ALL} shows the same comparison for the two IMR models. The NR surrogate generally agrees very well across the parameter space, although it can struggle with the largest timing differences (see Figure \ref{fig:q1_cases}). The EOB-based SEOBNRv5HM model's performance varies significantly with the mode and configuration. For the $(4, 4)$ mode, SEOBNRv5HM struggles with smaller timing differences but excels at the larger differences in equal-mass cases. Table \ref{qall_table} quantifies these results using the median of the absolute timing difference between each model and NR for all cases.

\section{Discussion}
The time at which a gravitational wave mode reaches its peak amplitude is a key strong-field feature for decoding the physics of mergers. While the time differences between modes are small compared to the full waveform duration, accurately modeling them is critical for building subdominant mode models for next-generation detectors. The strong-field dynamics governing these peak times have led most waveform models to rely heavily on NR, either by directly interpolating simulations or by calibrating analytical models to NR catalogs. In this paper, we present two semi-analytical methods to predict these peak time differences without such calibration: one based on the Backwards-One-Body (BOB) model and another based on a timelike geodesic in the remnant black hole spacetime. Both models require only the remnant mass, spin, and the frequency at the time of peak strain amplitude for each mode as input from NR. NR frequency information is already a necessary ingredient in IMR modeling approaches, such as in \cite{Mahesh:2025}, for connecting inspiral and merger waveforms. 

We demonstrate that this same NR information can be leveraged to determine a different strong-field physical parameter, the relative time between the peaks of higher harmonics. The semi-analytical approaches introduced here help reduce the heavy reliance on NR for strong-field predictions. Due to the large parameter space that needs to be modeled, and the limited size of NR catalogs, waveform models that heavily calibrate to or interpolate among available NR results can yield inaccurate predictions when applied to new cases, even when those cases lie within the region of parameter space where they were tuned, as seen in Figs. \ref{fig:q1_cases} and \ref{figure:OTHER_ALL}. 

The BOB method's success in predicting these timings further validates its frequency evolution model through the merger-ringdown. The geodesic model's success suggests that despite the merger's non-linearity, mode peak excitation is effectively approximated by test-particle dynamics in the remnant Kerr spacetime. Furthermore, the simple linear relationships between the characteristic radii we calculate and the minimum of the effective potentials for scalar and tensor radiation in the background of the remnant Kerr spacetime suggest that we may be able to more directly model the characteristic radii in future work. Ultimately, these findings imply that while the merger itself is non-linear, the relative peak timings of the observed radiation are well described by linear physics on the background of the remnant black hole.

We find that both our models accurately predict the timing differences across a wide range of the parameter space for quasi-circular, non-precessing systems and for $l=|m|$ modes up to $l=8$. They perform particularly well for equal-mass cases, which exhibit the largest timing differences. Both models perform worst for the $(3, 3)$ mode, but show consistently excellent results for higher modes. In comparison, we find the SEOBNRv5HM model performance varies, while NRHybSur3dq8 agrees very well with NR except in certain cases. Notably, for the equal-mass $(5, 5)$ mode, where the timing differences are large, both our models significantly outperform the two IMR models. This demonstrates the potential for semi-analytical methods to capture complex merger physics.

This work was constrained to equatorial orbits and non-precessing systems, limiting our analysis to $l=|m|$ modes. Future work will extend these methods to non-equatorial orbits to model precessing binaries and a wider variety of modes.

\begin{acknowledgments}
We thank Matthew Cerep and Siddarth Mahesh for useful discussions. AK and STM were supported in part by NASA grant 22-LPS22-0022 and 24-2024EPSCoR-0010. This research was made possible by the NASA West Virginia Space Grant Consortium, Grant \# 80NSSC20M0055. The authors acknowledge the computational resources provided by the WVU Research Computing Thorny Flat HPC cluster, which is funded in part by NSF OAC-1726534.
\end{acknowledgments}

\appendix
\section{Excluded SXS Simulations}
\begin{table}[htpb]
\begin{tabular}{l@{\hskip 2em}l}
\toprule
\textbf{SXS ID} & \textbf{Modes Removed} \\
\midrule
0160 & (3,3), (5,5), (6,6), (7,7), $(8,8)^*$ \\
0172 & (3,3), (5,5), (6,6), $(7,7)^*$, (8,8) \\
0176 & (3,3), (5,5), (6,6), (7,7), (8,8) \\
0178 & (3,3), (5,5), (6,6), (7,7), (8,8) \\
0185 & (7,7), (8,8) \\
0192 & (7,7), (8,8) \\
0195 & (6,6), (8,8) \\
0198 & (5,5), (6,6), (7,7), (8,8) \\
0201 & (5,5), (6,6), (7,7), (8,8) \\
0202 & (5,5), (6,6), (7,7), (8,8) \\
0203 & (5,5), (6,6), (7,7), (8,8) \\
0204 & (3,3), (4,4), (5,5), (6,6), (7,7), (8,8) \\
0205 & (6,6), (8,8) \\
0206 & (3,3), (4,4), (5,5), (7,7), (8,8) \\
0207 & (7,7), (8,8) \\
0208 & (3,3), (4,4), (5,5), (8,8) \\
0292 & (6,6), (7,7), (8,8) \\
0293 & (6,6), (7,7), (8,8) \\
0317 & (7,7), (8,8) \\
0329 & (3,3), (5,5), (7,7) \\
0377 & (7,7) \\
0389 & (3,3), (5,5), (7,7), (8,8) \\
0486 & (3,3), (7,7), (8,8) \\
0503 & (3,3), (8,8) \\
0610 & (6,6), (7,7), (8,8) \\
0611 & (7,7) \\
0613 & (7,7), (8,8) \\
0614 & (7,7) \\
1122 & (3,3), (4,4), (5,5), (6,6), (7,7), (8,8) \\
1123 & (3,3), (5,5), (6,6), (7,7), (8,8) \\
1132 & (3,3), (4,4), (5,5), (6,6), (7,7), (8,8) \\
1137 & (3,3), (5,5), (6,6), (7,7), (8,8) \\
1141 & (3,3), (5,5), (7,7) \\
1146 & (5,5), (8,8) \\
1148 & (7,7), (8,8) \\
1153 & $(3,3)^*$, (5,5), (7,7), (8,8) \\
1154 & (3,3), (5,5), (7,7), (8,8) \\
1155 & (3,3), (5,5), (7,7), (8,8) \\
1221 & (5,5), (7,7), (8,8) \\
1222 & (4,4), (5,5), (6,6), (7,7), (8,8) \\
1351 & (3,3), (7,7), (8,8) \\
1412 & (3,3), (5,5), (7,7), (8,8) \\
1413 & (3,3), (4,4), (5,5), (6,6), (7,7), (8,8) \\
1414 & (3,3), (5,5), (6,6), (7,7), (8,8) \\
1415 & (3,3), (4,4), (5,5), (6,6), (7,7), (8,8) \\
1416 & (3,3), (5,5), (8,8) \\
1417 & (3,3), (6,6), (7,7), (8,8) \\
1426 & (7,7), (8,8) \\
1429 & (7,7), (8,8) \\
1430 & (7,7), (8,8) \\
1432 & (7,7), (8,8) \\
1437 & (7,7), (8,8) \\
1440 & (7,7), (8,8) \\
1463 & (7,7), (8,8) \\
1475 & (3,3), (5,5), (7,7) \\
\end{tabular}
\caption{List of quasicircular and non-precessing SXS simulations and the corresponding modes that were removed for low quality, as determined by the filtering criteria defined in Section~\ref{subsec:Error}. We do not include a large number of simulations in which only the $(8,8)$ mode was removed as well as any simulations in which lower resolution data was not available. Asterisks indicate modes that were not caught by our automatic filters but were removed due to their apparent poor quality after a visual inspection.}
\label{tab:bad_sims_1}
\end{table}
\begin{table}[]
\begin{tabular}{l@{\hskip 2em}l}
\toprule
\textbf{SXS ID} & \textbf{Modes Removed} \\
\midrule
1487 & (3,3) \\
1490 & (7,7), (8,8) \\
1494 & (4,4), (5,5), (6,6), (7,7), (8,8) \\
1500 & (7,7) \\
2086 & (3,3), (5,5), (7,7) \\
2089 & (3,3), (5,5), (7,7), (8,8) \\
2102 & (3,3), (5,5), (7,7) \\
2190 & (7,7), (8,8) \\
2325 & (3,3), (5,5), (6,6), (7,7), (8,8) \\
2326 & (3,3), (5,5), (6,6), (7,7), (8,8) \\
2375 & $(3,3)^*$, (5,5), (6,6), (7,7), (8,8) \\
2376 & (3,3), (5,5), (6,6), (7,7), (8,8) \\
2377 & (3,3), (5,5), (6,6), (7,7), (8,8) \\
2378 & (3,3), (5,5), (6,6), (7,7), (8,8) \\
2418 & (3,3), (5,5), (7,7) \\
2419 & (3,3), (5,5), (7,7) \\
2420 & (3,3), (5,5), (7,7) \\
2421 & (3,3), (5,5), (7,7), (8,8) \\
2422 & (3,3), (5,5), (7,7), (8,8) \\
2423 & (3,3), (5,5), (7,7), (8,8) \\
2486 & (7,7), (8,8) \\
2494 & (4,4), (5,5), (6,6), (7,7), (8,8) \\
2496 & (3,3), (5,5), (7,7), (8,8) \\
2500 & (3,3), (5,5), (7,7), (8,8) \\
2505 & (3,3), (5,5), (7,7) \\
2509 & (3,3), (5,5), (6,6), (7,7), (8,8) \\
2512 & (3,3), (5,5), (7,7), (8,8) \\
2669 & (7,7), (8,8) \\
2677 & (7,7), (8,8) \\
2696 & (7,7), (8,8) \\
2700 & (6,6), (7,7) \\
2786 & (7,7), $(8,8)^*$ \\
3518 & (3,3), (5,5), (7,7), (8,8) \\
3519 & (6,6), (8,8) \\
3553 & (6,6), (8,8) \\
3582 & (6,6) \\
3617 & (3,3), (5,5), (6,6), (7,7), (8,8) \\
3624 & (3,3), (5,5), (7,7), (8,8) \\
3625 & (3,3), (5,5), (7,7), (8,8) \\
3626 & (5,5), (7,7) \\
3627 & (3,3), (4,4), (5,5), (7,7), (8,8) \\
3628 & (3,3), (5,5), (6,6), (7,7), (8,8) \\
3629 & (3,3), (5,5), (7,7), (8,8) \\
3632 & (3,3), (5,5), (7,7) \\
3633 & (3,3), (5,5), (7,7) \\
3634 & (3,3), (5,5), (7,7), (8,8) \\
3864 & (3,3), (5,5), (7,7) \\
3895 & (3,3), (5,5), (7,7) \\
3897 & (3,3), (5,5), (7,7), (8,8) \\
3917 & (3,3), (7,7), (8,8) \\
3976 & (3,3), (5,5), (7,7), (8,8) \\
3978 & (3,3), (5,5), (7,7), (8,8) \\
3979 & (7,7), (8,8) \\
3980 & (7,7) \\
4072 & (3,3), (5,5), (7,7) \\
4236 & (7,7), (8,8) \\
4434 & (3,3), (5,5), (7,7) \\
\end{tabular}
\caption*{TABLE \ref{tab:bad_sims_1} \textit{(Continued).}}
\label{tab:bad_sims_2}
\end{table}

\clearpage
\bibliography{citations}

\begin{thebibliography}{39}%
\makeatletter
\providecommand \@ifxundefined [1]{%
 \@ifx{#1\undefined}
}%
\providecommand \@ifnum [1]{%
 \ifnum #1\expandafter \@firstoftwo
 \else \expandafter \@secondoftwo
 \fi
}%
\providecommand \@ifx [1]{%
 \ifx #1\expandafter \@firstoftwo
 \else \expandafter \@secondoftwo
 \fi
}%
\providecommand \natexlab [1]{#1}%
\providecommand \enquote  [1]{``#1''}%
\providecommand \bibnamefont  [1]{#1}%
\providecommand \bibfnamefont [1]{#1}%
\providecommand \citenamefont [1]{#1}%
\providecommand \href@noop [0]{\@secondoftwo}%
\providecommand \href [0]{\begingroup \@sanitize@url \@href}%
\providecommand \@href[1]{\@@startlink{#1}\@@href}%
\providecommand \@@href[1]{\endgroup#1\@@endlink}%
\providecommand \@sanitize@url [0]{\catcode `\\12\catcode `\$12\catcode `\&12\catcode `\#12\catcode `\^12\catcode `\_12\catcode `\%12\relax}%
\providecommand \@@startlink[1]{}%
\providecommand \@@endlink[0]{}%
\providecommand \url  [0]{\begingroup\@sanitize@url \@url }%
\providecommand \@url [1]{\endgroup\@href {#1}{\urlprefix }}%
\providecommand \urlprefix  [0]{URL }%
\providecommand \Eprint [0]{\href }%
\providecommand \doibase [0]{https://doi.org/}%
\providecommand \selectlanguage [0]{\@gobble}%
\providecommand \bibinfo  [0]{\@secondoftwo}%
\providecommand \bibfield  [0]{\@secondoftwo}%
\providecommand \translation [1]{[#1]}%
\providecommand \BibitemOpen [0]{}%
\providecommand \bibitemStop [0]{}%
\providecommand \bibitemNoStop [0]{.\EOS\space}%
\providecommand \EOS [0]{\spacefactor3000\relax}%
\providecommand \BibitemShut  [1]{\csname bibitem#1\endcsname}%
\let\auto@bib@innerbib\@empty
\bibitem [{\citenamefont {Abbott}\ \emph {et~al.}(2021{\natexlab{a}})\citenamefont {Abbott}, \citenamefont {Abbott}, \citenamefont {Abraham}, \citenamefont {Acernese}, \citenamefont {Ackley}, \citenamefont {Adams}, \citenamefont {Adams}, \citenamefont {Adhikari}, \citenamefont {Adya}, \citenamefont {Affeldt} \emph {et~al.}}]{evol_cite1}%
  \BibitemOpen
  \bibfield  {author} {\bibinfo {author} {\bibfnamefont {R.}~\bibnamefont {Abbott}}, \bibinfo {author} {\bibfnamefont {T.}~\bibnamefont {Abbott}}, \bibinfo {author} {\bibfnamefont {S.}~\bibnamefont {Abraham}}, \bibinfo {author} {\bibfnamefont {F.}~\bibnamefont {Acernese}}, \bibinfo {author} {\bibfnamefont {K.}~\bibnamefont {Ackley}}, \bibinfo {author} {\bibfnamefont {A.}~\bibnamefont {Adams}}, \bibinfo {author} {\bibfnamefont {C.}~\bibnamefont {Adams}}, \bibinfo {author} {\bibfnamefont {R.}~\bibnamefont {Adhikari}}, \bibinfo {author} {\bibfnamefont {V.}~\bibnamefont {Adya}}, \bibinfo {author} {\bibfnamefont {C.}~\bibnamefont {Affeldt}}, \emph {et~al.},\ }\href@noop {} {\bibfield  {journal} {\bibinfo  {journal} {Phys. Rev. X}\ }\textbf {\bibinfo {volume} {11}},\ \bibinfo {pages} {021053} (\bibinfo {year} {2021}{\natexlab{a}})}\BibitemShut {NoStop}%
\bibitem [{\citenamefont {Abbott}\ \emph {et~al.}(2023)\citenamefont {Abbott}, \citenamefont {Abbott}, \citenamefont {Acernese}, \citenamefont {Ackley}, \citenamefont {Adams}, \citenamefont {Adhikari}, \citenamefont {Adhikari}, \citenamefont {Adya}, \citenamefont {Affeldt}, \citenamefont {Agarwal} \emph {et~al.}}]{evol_cite2}%
  \BibitemOpen
  \bibfield  {author} {\bibinfo {author} {\bibfnamefont {R.}~\bibnamefont {Abbott}}, \bibinfo {author} {\bibfnamefont {T.}~\bibnamefont {Abbott}}, \bibinfo {author} {\bibfnamefont {F.}~\bibnamefont {Acernese}}, \bibinfo {author} {\bibfnamefont {K.}~\bibnamefont {Ackley}}, \bibinfo {author} {\bibfnamefont {C.}~\bibnamefont {Adams}}, \bibinfo {author} {\bibfnamefont {N.}~\bibnamefont {Adhikari}}, \bibinfo {author} {\bibfnamefont {R.}~\bibnamefont {Adhikari}}, \bibinfo {author} {\bibfnamefont {V.}~\bibnamefont {Adya}}, \bibinfo {author} {\bibfnamefont {C.}~\bibnamefont {Affeldt}}, \bibinfo {author} {\bibfnamefont {D.}~\bibnamefont {Agarwal}}, \emph {et~al.},\ }\href@noop {} {\bibfield  {journal} {\bibinfo  {journal} {Phys. Rev. X}\ }\textbf {\bibinfo {volume} {13}},\ \bibinfo {pages} {041039} (\bibinfo {year} {2023})}\BibitemShut {NoStop}%
\bibitem [{\citenamefont {Abbott}\ \emph {et~al.}(2021{\natexlab{b}})\citenamefont {Abbott}, \citenamefont {Abbott}, \citenamefont {Abraham}, \citenamefont {Acernese}, \citenamefont {Ackley}, \citenamefont {Adams}, \citenamefont {Adams}, \citenamefont {Adhikari}, \citenamefont {Adya}, \citenamefont {Affeldt} \emph {et~al.}}]{strong_field_GR_test}%
  \BibitemOpen
  \bibfield  {author} {\bibinfo {author} {\bibfnamefont {R.}~\bibnamefont {Abbott}}, \bibinfo {author} {\bibfnamefont {T.}~\bibnamefont {Abbott}}, \bibinfo {author} {\bibfnamefont {S.}~\bibnamefont {Abraham}}, \bibinfo {author} {\bibfnamefont {F.}~\bibnamefont {Acernese}}, \bibinfo {author} {\bibfnamefont {K.}~\bibnamefont {Ackley}}, \bibinfo {author} {\bibfnamefont {A.}~\bibnamefont {Adams}}, \bibinfo {author} {\bibfnamefont {C.}~\bibnamefont {Adams}}, \bibinfo {author} {\bibfnamefont {R.~X.}\ \bibnamefont {Adhikari}}, \bibinfo {author} {\bibfnamefont {V.}~\bibnamefont {Adya}}, \bibinfo {author} {\bibfnamefont {C.}~\bibnamefont {Affeldt}}, \emph {et~al.},\ }\href@noop {} {\bibfield  {journal} {\bibinfo  {journal} {Phys. Rev. D}\ }\textbf {\bibinfo {volume} {103}},\ \bibinfo {pages} {122002} (\bibinfo {year} {2021}{\natexlab{b}})}\BibitemShut {NoStop}%
\bibitem [{\citenamefont {Amaro-Seoane}\ \emph {et~al.}(2017)\citenamefont {Amaro-Seoane}, \citenamefont {Audley}, \citenamefont {Babak}, \citenamefont {Baker}, \citenamefont {Barausse}, \citenamefont {Bender}, \citenamefont {Berti}, \citenamefont {Binetruy}, \citenamefont {Born}, \citenamefont {Bortoluzzi} \emph {et~al.}}]{LISA}%
  \BibitemOpen
  \bibfield  {author} {\bibinfo {author} {\bibfnamefont {P.}~\bibnamefont {Amaro-Seoane}}, \bibinfo {author} {\bibfnamefont {H.}~\bibnamefont {Audley}}, \bibinfo {author} {\bibfnamefont {S.}~\bibnamefont {Babak}}, \bibinfo {author} {\bibfnamefont {J.}~\bibnamefont {Baker}}, \bibinfo {author} {\bibfnamefont {E.}~\bibnamefont {Barausse}}, \bibinfo {author} {\bibfnamefont {P.}~\bibnamefont {Bender}}, \bibinfo {author} {\bibfnamefont {E.}~\bibnamefont {Berti}}, \bibinfo {author} {\bibfnamefont {P.}~\bibnamefont {Binetruy}}, \bibinfo {author} {\bibfnamefont {M.}~\bibnamefont {Born}}, \bibinfo {author} {\bibfnamefont {D.}~\bibnamefont {Bortoluzzi}}, \emph {et~al.},\ }\href@noop {} {\bibinfo {title} {Laser interferometer space antenna}} (\bibinfo {year} {2017}),\ \Eprint {https://arxiv.org/abs/1702.00786} {arXiv:1702.00786 [astro-ph.IM]} \BibitemShut {NoStop}%
\bibitem [{\citenamefont {Colpi}\ \emph {et~al.}(2024)\citenamefont {Colpi}, \citenamefont {Danzmann}, \citenamefont {Hewitson}, \citenamefont {Holley-Bockelmann}, \citenamefont {Jetzer}, \citenamefont {Nelemans}, \citenamefont {Petiteau}, \citenamefont {Shoemaker}, \citenamefont {Sopuerta}, \citenamefont {Stebbins} \emph {et~al.}}]{LISA_cite2}%
  \BibitemOpen
  \bibfield  {author} {\bibinfo {author} {\bibfnamefont {M.}~\bibnamefont {Colpi}}, \bibinfo {author} {\bibfnamefont {K.}~\bibnamefont {Danzmann}}, \bibinfo {author} {\bibfnamefont {M.}~\bibnamefont {Hewitson}}, \bibinfo {author} {\bibfnamefont {K.}~\bibnamefont {Holley-Bockelmann}}, \bibinfo {author} {\bibfnamefont {P.}~\bibnamefont {Jetzer}}, \bibinfo {author} {\bibfnamefont {G.}~\bibnamefont {Nelemans}}, \bibinfo {author} {\bibfnamefont {A.}~\bibnamefont {Petiteau}}, \bibinfo {author} {\bibfnamefont {D.}~\bibnamefont {Shoemaker}}, \bibinfo {author} {\bibfnamefont {C.}~\bibnamefont {Sopuerta}}, \bibinfo {author} {\bibfnamefont {R.}~\bibnamefont {Stebbins}}, \emph {et~al.},\ }\href@noop {} {\bibfield  {journal} {\bibinfo  {journal} {arXiv preprint arXiv:2402.07571}\ } (\bibinfo {year} {2024})}\BibitemShut {NoStop}%
\bibitem [{\citenamefont {Reitze}\ \emph {et~al.}(2019)\citenamefont {Reitze} \emph {et~al.}}]{CE_cite1}%
  \BibitemOpen
  \bibfield  {author} {\bibinfo {author} {\bibfnamefont {D.}~\bibnamefont {Reitze}} \emph {et~al.},\ }\href@noop {} {\bibfield  {journal} {\bibinfo  {journal} {Bull. Am. Astron. Soc.}\ }\textbf {\bibinfo {volume} {51}},\ \bibinfo {pages} {035} (\bibinfo {year} {2019})},\ \Eprint {https://arxiv.org/abs/1907.04833} {arXiv:1907.04833 [astro-ph.IM]} \BibitemShut {NoStop}%
\bibitem [{\citenamefont {Evans}\ \emph {et~al.}(2021)\citenamefont {Evans}, \citenamefont {Adhikari}, \citenamefont {Afle}, \citenamefont {Ballmer}, \citenamefont {Biscoveanu}, \citenamefont {Borhanian}, \citenamefont {Brown}, \citenamefont {Chen}, \citenamefont {Eisenstein}, \citenamefont {Gruson} \emph {et~al.}}]{CE_cite2}%
  \BibitemOpen
  \bibfield  {author} {\bibinfo {author} {\bibfnamefont {M.}~\bibnamefont {Evans}}, \bibinfo {author} {\bibfnamefont {R.~X.}\ \bibnamefont {Adhikari}}, \bibinfo {author} {\bibfnamefont {C.}~\bibnamefont {Afle}}, \bibinfo {author} {\bibfnamefont {S.~W.}\ \bibnamefont {Ballmer}}, \bibinfo {author} {\bibfnamefont {S.}~\bibnamefont {Biscoveanu}}, \bibinfo {author} {\bibfnamefont {S.}~\bibnamefont {Borhanian}}, \bibinfo {author} {\bibfnamefont {D.~A.}\ \bibnamefont {Brown}}, \bibinfo {author} {\bibfnamefont {Y.}~\bibnamefont {Chen}}, \bibinfo {author} {\bibfnamefont {R.}~\bibnamefont {Eisenstein}}, \bibinfo {author} {\bibfnamefont {A.}~\bibnamefont {Gruson}}, \emph {et~al.},\ }\href@noop {} {\bibfield  {journal} {\bibinfo  {journal} {arXiv preprint arXiv:2109.09882}\ } (\bibinfo {year} {2021})}\BibitemShut {NoStop}%
\bibitem [{\citenamefont {Buikema}\ \emph {et~al.}(2020)\citenamefont {Buikema}, \citenamefont {Cahillane}, \citenamefont {Mansell}, \citenamefont {Blair}, \citenamefont {Abbott}, \citenamefont {Adams}, \citenamefont {Adhikari}, \citenamefont {Ananyeva}, \citenamefont {Appert}, \citenamefont {Arai} \emph {et~al.}}]{LIGO_cite1}%
  \BibitemOpen
  \bibfield  {author} {\bibinfo {author} {\bibfnamefont {A.}~\bibnamefont {Buikema}}, \bibinfo {author} {\bibfnamefont {C.}~\bibnamefont {Cahillane}}, \bibinfo {author} {\bibfnamefont {G.}~\bibnamefont {Mansell}}, \bibinfo {author} {\bibfnamefont {C.}~\bibnamefont {Blair}}, \bibinfo {author} {\bibfnamefont {R.}~\bibnamefont {Abbott}}, \bibinfo {author} {\bibfnamefont {C.}~\bibnamefont {Adams}}, \bibinfo {author} {\bibfnamefont {R.}~\bibnamefont {Adhikari}}, \bibinfo {author} {\bibfnamefont {A.}~\bibnamefont {Ananyeva}}, \bibinfo {author} {\bibfnamefont {S.}~\bibnamefont {Appert}}, \bibinfo {author} {\bibfnamefont {K.}~\bibnamefont {Arai}}, \emph {et~al.},\ }\href@noop {} {\bibfield  {journal} {\bibinfo  {journal} {Phys. Rev. D}\ }\textbf {\bibinfo {volume} {102}},\ \bibinfo {pages} {062003} (\bibinfo {year} {2020})}\BibitemShut {NoStop}%
\bibitem [{\citenamefont {Aasi}\ \emph {et~al.}(2015)\citenamefont {Aasi}, \citenamefont {Abbott}, \citenamefont {Abbott}, \citenamefont {Abbott}, \citenamefont {Abernathy}, \citenamefont {Ackley}, \citenamefont {Adams}, \citenamefont {Adams}, \citenamefont {Addesso}, \citenamefont {Adhikari} \emph {et~al.}}]{LIGO_cite2}%
  \BibitemOpen
  \bibfield  {author} {\bibinfo {author} {\bibfnamefont {J.}~\bibnamefont {Aasi}}, \bibinfo {author} {\bibfnamefont {B.}~\bibnamefont {Abbott}}, \bibinfo {author} {\bibfnamefont {R.}~\bibnamefont {Abbott}}, \bibinfo {author} {\bibfnamefont {T.}~\bibnamefont {Abbott}}, \bibinfo {author} {\bibfnamefont {M.}~\bibnamefont {Abernathy}}, \bibinfo {author} {\bibfnamefont {K.}~\bibnamefont {Ackley}}, \bibinfo {author} {\bibfnamefont {C.}~\bibnamefont {Adams}}, \bibinfo {author} {\bibfnamefont {T.}~\bibnamefont {Adams}}, \bibinfo {author} {\bibfnamefont {P.}~\bibnamefont {Addesso}}, \bibinfo {author} {\bibfnamefont {R.}~\bibnamefont {Adhikari}}, \emph {et~al.},\ }\href@noop {} {\bibfield  {journal} {\bibinfo  {journal} {Class. Quantum Grav.}\ }\textbf {\bibinfo {volume} {32}},\ \bibinfo {pages} {074001} (\bibinfo {year} {2015})}\BibitemShut {NoStop}%
\bibitem [{\citenamefont {Pitte}\ \emph {et~al.}(2023)\citenamefont {Pitte}, \citenamefont {Baghi}, \citenamefont {Marsat}, \citenamefont {Besan{\c{c}}on},\ and\ \citenamefont {Petiteau}}]{higher_mode_cite1}%
  \BibitemOpen
  \bibfield  {author} {\bibinfo {author} {\bibfnamefont {C.}~\bibnamefont {Pitte}}, \bibinfo {author} {\bibfnamefont {Q.}~\bibnamefont {Baghi}}, \bibinfo {author} {\bibfnamefont {S.}~\bibnamefont {Marsat}}, \bibinfo {author} {\bibfnamefont {M.}~\bibnamefont {Besan{\c{c}}on}},\ and\ \bibinfo {author} {\bibfnamefont {A.}~\bibnamefont {Petiteau}},\ }\href@noop {} {\bibfield  {journal} {\bibinfo  {journal} {Phys. Rev. D}\ }\textbf {\bibinfo {volume} {108}},\ \bibinfo {pages} {044053} (\bibinfo {year} {2023})}\BibitemShut {NoStop}%
\bibitem [{\citenamefont {Gong}\ \emph {et~al.}(2023)\citenamefont {Gong}, \citenamefont {Cao}, \citenamefont {Zhao},\ and\ \citenamefont {Shao}}]{higher_mode_cite2}%
  \BibitemOpen
  \bibfield  {author} {\bibinfo {author} {\bibfnamefont {Y.}~\bibnamefont {Gong}}, \bibinfo {author} {\bibfnamefont {Z.}~\bibnamefont {Cao}}, \bibinfo {author} {\bibfnamefont {J.}~\bibnamefont {Zhao}},\ and\ \bibinfo {author} {\bibfnamefont {L.}~\bibnamefont {Shao}},\ }\href@noop {} {\bibfield  {journal} {\bibinfo  {journal} {Physical Review D}\ }\textbf {\bibinfo {volume} {108}},\ \bibinfo {pages} {064046} (\bibinfo {year} {2023})}\BibitemShut {NoStop}%
\bibitem [{\citenamefont {Frattale~Mascioli}\ \emph {et~al.}(2025)\citenamefont {Frattale~Mascioli}, \citenamefont {Crescimbeni}, \citenamefont {Pacilio}, \citenamefont {Pani},\ and\ \citenamefont {Pannarale}}]{higher_mode_degen1}%
  \BibitemOpen
  \bibfield  {author} {\bibinfo {author} {\bibfnamefont {A.}~\bibnamefont {Frattale~Mascioli}}, \bibinfo {author} {\bibfnamefont {F.}~\bibnamefont {Crescimbeni}}, \bibinfo {author} {\bibfnamefont {C.}~\bibnamefont {Pacilio}}, \bibinfo {author} {\bibfnamefont {P.}~\bibnamefont {Pani}},\ and\ \bibinfo {author} {\bibfnamefont {F.}~\bibnamefont {Pannarale}},\ }\href@noop {} {\bibfield  {journal} {\bibinfo  {journal} {arXiv e-prints}\ } (\bibinfo {year} {2025})}\BibitemShut {NoStop}%
\bibitem [{\citenamefont {London}\ \emph {et~al.}(2018)\citenamefont {London}, \citenamefont {Khan}, \citenamefont {Fauchon-Jones}, \citenamefont {Garc{\'\i}a}, \citenamefont {Hannam}, \citenamefont {Husa}, \citenamefont {Jim{\'e}nez-Forteza}, \citenamefont {Kalaghatgi}, \citenamefont {Ohme},\ and\ \citenamefont {Pannarale}}]{higher_mode_degen2}%
  \BibitemOpen
  \bibfield  {author} {\bibinfo {author} {\bibfnamefont {L.}~\bibnamefont {London}}, \bibinfo {author} {\bibfnamefont {S.}~\bibnamefont {Khan}}, \bibinfo {author} {\bibfnamefont {E.}~\bibnamefont {Fauchon-Jones}}, \bibinfo {author} {\bibfnamefont {C.}~\bibnamefont {Garc{\'\i}a}}, \bibinfo {author} {\bibfnamefont {M.}~\bibnamefont {Hannam}}, \bibinfo {author} {\bibfnamefont {S.}~\bibnamefont {Husa}}, \bibinfo {author} {\bibfnamefont {X.}~\bibnamefont {Jim{\'e}nez-Forteza}}, \bibinfo {author} {\bibfnamefont {C.}~\bibnamefont {Kalaghatgi}}, \bibinfo {author} {\bibfnamefont {F.}~\bibnamefont {Ohme}},\ and\ \bibinfo {author} {\bibfnamefont {F.}~\bibnamefont {Pannarale}},\ }\href@noop {} {\bibfield  {journal} {\bibinfo  {journal} {Physical review letters}\ }\textbf {\bibinfo {volume} {120}},\ \bibinfo {pages} {161102} (\bibinfo {year} {2018})}\BibitemShut {NoStop}%
\bibitem [{\citenamefont {Borhanian}\ \emph {et~al.}(2020)\citenamefont {Borhanian}, \citenamefont {Dhani}, \citenamefont {Gupta}, \citenamefont {Arun},\ and\ \citenamefont {Sathyaprakash}}]{higher_mode_distance1}%
  \BibitemOpen
  \bibfield  {author} {\bibinfo {author} {\bibfnamefont {S.}~\bibnamefont {Borhanian}}, \bibinfo {author} {\bibfnamefont {A.}~\bibnamefont {Dhani}}, \bibinfo {author} {\bibfnamefont {A.}~\bibnamefont {Gupta}}, \bibinfo {author} {\bibfnamefont {K.}~\bibnamefont {Arun}},\ and\ \bibinfo {author} {\bibfnamefont {B.}~\bibnamefont {Sathyaprakash}},\ }\href@noop {} {\bibfield  {journal} {\bibinfo  {journal} {The Astrophysical Journal Letters}\ }\textbf {\bibinfo {volume} {905}},\ \bibinfo {pages} {L28} (\bibinfo {year} {2020})}\BibitemShut {NoStop}%
\bibitem [{\citenamefont {Puecher}\ \emph {et~al.}(2022)\citenamefont {Puecher}, \citenamefont {Kalaghatgi}, \citenamefont {Roy}, \citenamefont {Setyawati}, \citenamefont {Gupta}, \citenamefont {Sathyaprakash},\ and\ \citenamefont {Van Den~Broeck}}]{higher_mode_grtest1}%
  \BibitemOpen
  \bibfield  {author} {\bibinfo {author} {\bibfnamefont {A.}~\bibnamefont {Puecher}}, \bibinfo {author} {\bibfnamefont {C.}~\bibnamefont {Kalaghatgi}}, \bibinfo {author} {\bibfnamefont {S.}~\bibnamefont {Roy}}, \bibinfo {author} {\bibfnamefont {Y.}~\bibnamefont {Setyawati}}, \bibinfo {author} {\bibfnamefont {I.}~\bibnamefont {Gupta}}, \bibinfo {author} {\bibfnamefont {B.}~\bibnamefont {Sathyaprakash}},\ and\ \bibinfo {author} {\bibfnamefont {C.}~\bibnamefont {Van Den~Broeck}},\ }\href@noop {} {\bibfield  {journal} {\bibinfo  {journal} {Physical Review D}\ }\textbf {\bibinfo {volume} {106}},\ \bibinfo {pages} {082003} (\bibinfo {year} {2022})}\BibitemShut {NoStop}%
\bibitem [{\citenamefont {Mehta}\ \emph {et~al.}(2023)\citenamefont {Mehta}, \citenamefont {Buonanno}, \citenamefont {Cotesta}, \citenamefont {Ghosh}, \citenamefont {Sennett},\ and\ \citenamefont {Steinhoff}}]{higher_mode_grtest2}%
  \BibitemOpen
  \bibfield  {author} {\bibinfo {author} {\bibfnamefont {A.~K.}\ \bibnamefont {Mehta}}, \bibinfo {author} {\bibfnamefont {A.}~\bibnamefont {Buonanno}}, \bibinfo {author} {\bibfnamefont {R.}~\bibnamefont {Cotesta}}, \bibinfo {author} {\bibfnamefont {A.}~\bibnamefont {Ghosh}}, \bibinfo {author} {\bibfnamefont {N.}~\bibnamefont {Sennett}},\ and\ \bibinfo {author} {\bibfnamefont {J.}~\bibnamefont {Steinhoff}},\ }\href@noop {} {\bibfield  {journal} {\bibinfo  {journal} {Physical Review D}\ }\textbf {\bibinfo {volume} {107}},\ \bibinfo {pages} {044020} (\bibinfo {year} {2023})}\BibitemShut {NoStop}%
\bibitem [{\citenamefont {Islam}(2021)}]{higher_mode_grtest3}%
  \BibitemOpen
  \bibfield  {author} {\bibinfo {author} {\bibfnamefont {T.}~\bibnamefont {Islam}},\ }\href@noop {} {\bibfield  {journal} {\bibinfo  {journal} {arXiv preprint arXiv:2111.00111}\ } (\bibinfo {year} {2021})}\BibitemShut {NoStop}%
\bibitem [{\citenamefont {Varma}\ \emph {et~al.}(2019)\citenamefont {Varma}, \citenamefont {Field}, \citenamefont {Scheel}, \citenamefont {Blackman}, \citenamefont {Kidder},\ and\ \citenamefont {Pfeiffer}}]{surr_cite1}%
  \BibitemOpen
  \bibfield  {author} {\bibinfo {author} {\bibfnamefont {V.}~\bibnamefont {Varma}}, \bibinfo {author} {\bibfnamefont {S.~E.}\ \bibnamefont {Field}}, \bibinfo {author} {\bibfnamefont {M.~A.}\ \bibnamefont {Scheel}}, \bibinfo {author} {\bibfnamefont {J.}~\bibnamefont {Blackman}}, \bibinfo {author} {\bibfnamefont {L.~E.}\ \bibnamefont {Kidder}},\ and\ \bibinfo {author} {\bibfnamefont {H.~P.}\ \bibnamefont {Pfeiffer}},\ }\href@noop {} {\bibfield  {journal} {\bibinfo  {journal} {Phys. Rev. D}\ }\textbf {\bibinfo {volume} {99}},\ \bibinfo {pages} {064045} (\bibinfo {year} {2019})}\BibitemShut {NoStop}%
\bibitem [{\citenamefont {Blackman}\ \emph {et~al.}(2015)\citenamefont {Blackman}, \citenamefont {Field}, \citenamefont {Galley}, \citenamefont {Szil{\'a}gyi}, \citenamefont {Scheel}, \citenamefont {Tiglio},\ and\ \citenamefont {Hemberger}}]{surr_cite2}%
  \BibitemOpen
  \bibfield  {author} {\bibinfo {author} {\bibfnamefont {J.}~\bibnamefont {Blackman}}, \bibinfo {author} {\bibfnamefont {S.~E.}\ \bibnamefont {Field}}, \bibinfo {author} {\bibfnamefont {C.~R.}\ \bibnamefont {Galley}}, \bibinfo {author} {\bibfnamefont {B.}~\bibnamefont {Szil{\'a}gyi}}, \bibinfo {author} {\bibfnamefont {M.~A.}\ \bibnamefont {Scheel}}, \bibinfo {author} {\bibfnamefont {M.}~\bibnamefont {Tiglio}},\ and\ \bibinfo {author} {\bibfnamefont {D.~A.}\ \bibnamefont {Hemberger}},\ }\href@noop {} {\bibfield  {journal} {\bibinfo  {journal} {Phys. Rev. Lett.}\ }\textbf {\bibinfo {volume} {115}},\ \bibinfo {pages} {121102} (\bibinfo {year} {2015})}\BibitemShut {NoStop}%
\bibitem [{\citenamefont {Buonanno}\ and\ \citenamefont {Damour}(1999)}]{EOB_cite1}%
  \BibitemOpen
  \bibfield  {author} {\bibinfo {author} {\bibfnamefont {A.}~\bibnamefont {Buonanno}}\ and\ \bibinfo {author} {\bibfnamefont {T.}~\bibnamefont {Damour}},\ }\bibfield  {journal} {\bibinfo  {journal} {Phys. Rev. D}\ }\textbf {\bibinfo {volume} {59}},\ \href {https://doi.org/10.1103/physrevd.59.084006} {10.1103/physrevd.59.084006} (\bibinfo {year} {1999})\BibitemShut {NoStop}%
\bibitem [{\citenamefont {Damour}(2001)}]{EOB_cite2}%
  \BibitemOpen
  \bibfield  {author} {\bibinfo {author} {\bibfnamefont {T.}~\bibnamefont {Damour}},\ }\href {http://dx.doi.org/10.1103/PhysRevD.64.124013} {\bibfield  {journal} {\bibinfo  {journal} {Phys. Rev. D}\ }\textbf {\bibinfo {volume} {64}} (\bibinfo {year} {2001})}\BibitemShut {NoStop}%
\bibitem [{\citenamefont {Boh{\'e}}\ \emph {et~al.}(2017)\citenamefont {Boh{\'e}}, \citenamefont {Shao}, \citenamefont {Taracchini}, \citenamefont {Buonanno}, \citenamefont {Babak}, \citenamefont {Harry}, \citenamefont {Hinder}, \citenamefont {Ossokine}, \citenamefont {P{\"u}rrer}, \citenamefont {Raymond} \emph {et~al.}}]{SEOBNR_cite1}%
  \BibitemOpen
  \bibfield  {author} {\bibinfo {author} {\bibfnamefont {A.}~\bibnamefont {Boh{\'e}}}, \bibinfo {author} {\bibfnamefont {L.}~\bibnamefont {Shao}}, \bibinfo {author} {\bibfnamefont {A.}~\bibnamefont {Taracchini}}, \bibinfo {author} {\bibfnamefont {A.}~\bibnamefont {Buonanno}}, \bibinfo {author} {\bibfnamefont {S.}~\bibnamefont {Babak}}, \bibinfo {author} {\bibfnamefont {I.~W.}\ \bibnamefont {Harry}}, \bibinfo {author} {\bibfnamefont {I.}~\bibnamefont {Hinder}}, \bibinfo {author} {\bibfnamefont {S.}~\bibnamefont {Ossokine}}, \bibinfo {author} {\bibfnamefont {M.}~\bibnamefont {P{\"u}rrer}}, \bibinfo {author} {\bibfnamefont {V.}~\bibnamefont {Raymond}}, \emph {et~al.},\ }\href@noop {} {\bibfield  {journal} {\bibinfo  {journal} {Phys. Rev. D}\ }\textbf {\bibinfo {volume} {95}},\ \bibinfo {pages} {044028} (\bibinfo {year} {2017})}\BibitemShut {NoStop}%
\bibitem [{\citenamefont {Pompili}\ \emph {et~al.}(2023)\citenamefont {Pompili}, \citenamefont {Buonanno}, \citenamefont {Estell{\'e}s}, \citenamefont {Khalil}, \citenamefont {van~de Meent}, \citenamefont {Mihaylov}, \citenamefont {Ossokine}, \citenamefont {P{\"u}rrer}, \citenamefont {Ramos-Buades}, \citenamefont {Mehta} \emph {et~al.}}]{SEOBNR_cite2}%
  \BibitemOpen
  \bibfield  {author} {\bibinfo {author} {\bibfnamefont {L.}~\bibnamefont {Pompili}}, \bibinfo {author} {\bibfnamefont {A.}~\bibnamefont {Buonanno}}, \bibinfo {author} {\bibfnamefont {H.}~\bibnamefont {Estell{\'e}s}}, \bibinfo {author} {\bibfnamefont {M.}~\bibnamefont {Khalil}}, \bibinfo {author} {\bibfnamefont {M.}~\bibnamefont {van~de Meent}}, \bibinfo {author} {\bibfnamefont {D.~P.}\ \bibnamefont {Mihaylov}}, \bibinfo {author} {\bibfnamefont {S.}~\bibnamefont {Ossokine}}, \bibinfo {author} {\bibfnamefont {M.}~\bibnamefont {P{\"u}rrer}}, \bibinfo {author} {\bibfnamefont {A.}~\bibnamefont {Ramos-Buades}}, \bibinfo {author} {\bibfnamefont {A.~K.}\ \bibnamefont {Mehta}}, \emph {et~al.},\ }\href@noop {} {\bibfield  {journal} {\bibinfo  {journal} {Phys. Rev. D}\ }\textbf {\bibinfo {volume} {108}},\ \bibinfo {pages} {124035} (\bibinfo {year} {2023})}\BibitemShut {NoStop}%
\bibitem [{\citenamefont {Nagar}\ \emph {et~al.}(2018)\citenamefont {Nagar}, \citenamefont {Bernuzzi}, \citenamefont {Del~Pozzo}, \citenamefont {Riemenschneider}, \citenamefont {Akcay}, \citenamefont {Carullo}, \citenamefont {Fleig}, \citenamefont {Babak}, \citenamefont {Tsang}, \citenamefont {Colleoni} \emph {et~al.}}]{TEOB_cite1}%
  \BibitemOpen
  \bibfield  {author} {\bibinfo {author} {\bibfnamefont {A.}~\bibnamefont {Nagar}}, \bibinfo {author} {\bibfnamefont {S.}~\bibnamefont {Bernuzzi}}, \bibinfo {author} {\bibfnamefont {W.}~\bibnamefont {Del~Pozzo}}, \bibinfo {author} {\bibfnamefont {G.}~\bibnamefont {Riemenschneider}}, \bibinfo {author} {\bibfnamefont {S.}~\bibnamefont {Akcay}}, \bibinfo {author} {\bibfnamefont {G.}~\bibnamefont {Carullo}}, \bibinfo {author} {\bibfnamefont {P.}~\bibnamefont {Fleig}}, \bibinfo {author} {\bibfnamefont {S.}~\bibnamefont {Babak}}, \bibinfo {author} {\bibfnamefont {K.~W.}\ \bibnamefont {Tsang}}, \bibinfo {author} {\bibfnamefont {M.}~\bibnamefont {Colleoni}}, \emph {et~al.},\ }\href@noop {} {\bibfield  {journal} {\bibinfo  {journal} {Phys. Rev. D}\ }\textbf {\bibinfo {volume} {98}},\ \bibinfo {pages} {104052} (\bibinfo {year} {2018})}\BibitemShut {NoStop}%
\bibitem [{\citenamefont {Nagar}\ \emph {et~al.}(2024)\citenamefont {Nagar}, \citenamefont {Gamba}, \citenamefont {Rettegno}, \citenamefont {Fantini},\ and\ \citenamefont {Bernuzzi}}]{TEOB_cite2}%
  \BibitemOpen
  \bibfield  {author} {\bibinfo {author} {\bibfnamefont {A.}~\bibnamefont {Nagar}}, \bibinfo {author} {\bibfnamefont {R.}~\bibnamefont {Gamba}}, \bibinfo {author} {\bibfnamefont {P.}~\bibnamefont {Rettegno}}, \bibinfo {author} {\bibfnamefont {V.}~\bibnamefont {Fantini}},\ and\ \bibinfo {author} {\bibfnamefont {S.}~\bibnamefont {Bernuzzi}},\ }\href {https://arxiv.org/abs/2404.05288} {\bibinfo {title} {Effective-one-body waveform model for non-circularized, planar, coalescing black hole binaries: the importance of radiation reaction}} (\bibinfo {year} {2024}),\ \Eprint {https://arxiv.org/abs/2404.05288} {arXiv:2404.05288 [gr-qc]} \BibitemShut {NoStop}%
\bibitem [{\citenamefont {{Blanchet}}(2014)}]{PN_cite1}%
  \BibitemOpen
  \bibfield  {author} {\bibinfo {author} {\bibfnamefont {L.}~\bibnamefont {{Blanchet}}},\ }\href@noop {} {\bibfield  {journal} {\bibinfo  {journal} {Living Rev. Relativity}\ }\textbf {\bibinfo {volume} {17}},\ \bibinfo {eid} {2} (\bibinfo {year} {2014})}\BibitemShut {NoStop}%
\bibitem [{\citenamefont {Blanchet}(2010)}]{PN_cite2}%
  \BibitemOpen
  \bibfield  {author} {\bibinfo {author} {\bibfnamefont {L.}~\bibnamefont {Blanchet}},\ }in\ \href@noop {} {\emph {\bibinfo {booktitle} {Mass and Motion in General Relativity}}}\ (\bibinfo  {publisher} {Springer},\ \bibinfo {year} {2010})\ pp.\ \bibinfo {pages} {125--166}\BibitemShut {NoStop}%
\bibitem [{\citenamefont {Sch{\"a}fer}(2010)}]{PN_cite3}%
  \BibitemOpen
  \bibfield  {author} {\bibinfo {author} {\bibfnamefont {G.}~\bibnamefont {Sch{\"a}fer}},\ }in\ \href@noop {} {\emph {\bibinfo {booktitle} {Mass and motion in General Relativity}}}\ (\bibinfo  {publisher} {Springer},\ \bibinfo {year} {2010})\ pp.\ \bibinfo {pages} {167--210}\BibitemShut {NoStop}%
\bibitem [{\citenamefont {McWilliams}(2019)}]{BOB}%
  \BibitemOpen
  \bibfield  {author} {\bibinfo {author} {\bibfnamefont {S.~T.}\ \bibnamefont {McWilliams}},\ }\href@noop {} {\bibfield  {journal} {\bibinfo  {journal} {Phys. Rev. Lett.}\ }\textbf {\bibinfo {volume} {122}},\ \bibinfo {pages} {191102} (\bibinfo {year} {2019})}\BibitemShut {NoStop}%
\bibitem [{\citenamefont {Mahesh}\ \emph {et~al.}(2025)\citenamefont {Mahesh}, \citenamefont {McWilliams},\ and\ \citenamefont {Etienne}}]{Mahesh:2025}%
  \BibitemOpen
  \bibfield  {author} {\bibinfo {author} {\bibfnamefont {S.}~\bibnamefont {Mahesh}}, \bibinfo {author} {\bibfnamefont {S.~T.}\ \bibnamefont {McWilliams}},\ and\ \bibinfo {author} {\bibfnamefont {Z.}~\bibnamefont {Etienne}},\ }\href@noop {} {\bibfield  {journal} {\bibinfo  {journal} {arXiv preprint arXiv:2508.20418}\ } (\bibinfo {year} {2025})}\BibitemShut {NoStop}%
\bibitem [{\citenamefont {Buonanno}\ \emph {et~al.}(2008)\citenamefont {Buonanno}, \citenamefont {Kidder},\ and\ \citenamefont {Lehner}}]{final_mass}%
  \BibitemOpen
  \bibfield  {author} {\bibinfo {author} {\bibfnamefont {A.}~\bibnamefont {Buonanno}}, \bibinfo {author} {\bibfnamefont {L.~E.}\ \bibnamefont {Kidder}},\ and\ \bibinfo {author} {\bibfnamefont {L.}~\bibnamefont {Lehner}},\ }\bibfield  {journal} {\bibinfo  {journal} {Phys. Rev. D}\ }\textbf {\bibinfo {volume} {77}},\ \href {https://doi.org/10.1103/physrevd.77.026004} {10.1103/physrevd.77.026004} (\bibinfo {year} {2008})\BibitemShut {NoStop}%
\bibitem [{\citenamefont {Damour}\ and\ \citenamefont {Nagar}(2007)}]{final_mass_eob}%
  \BibitemOpen
  \bibfield  {author} {\bibinfo {author} {\bibfnamefont {T.}~\bibnamefont {Damour}}\ and\ \bibinfo {author} {\bibfnamefont {A.}~\bibnamefont {Nagar}},\ }\bibfield  {journal} {\bibinfo  {journal} {Phys. Rev. D}\ }\textbf {\bibinfo {volume} {76}},\ \href {https://doi.org/10.1103/physrevd.76.044003} {10.1103/physrevd.76.044003} (\bibinfo {year} {2007})\BibitemShut {NoStop}%
\bibitem [{Note1()}]{Note1}%
  \BibitemOpen
  \bibinfo {note} {$n$ denotes the overtone number; we will assume that only the least damped fundamental overtone $n=0$ contributes, and denote modes as $(l, m)$ hereafter.}\BibitemShut {Stop}%
\bibitem [{\citenamefont {Bardeen}\ \emph {et~al.}(1972)\citenamefont {Bardeen}, \citenamefont {Press},\ and\ \citenamefont {Teukolsky}}]{bardeen}%
  \BibitemOpen
  \bibfield  {author} {\bibinfo {author} {\bibfnamefont {J.~M.}\ \bibnamefont {Bardeen}}, \bibinfo {author} {\bibfnamefont {W.~H.}\ \bibnamefont {Press}},\ and\ \bibinfo {author} {\bibfnamefont {S.~A.}\ \bibnamefont {Teukolsky}},\ }\href@noop {} {\bibfield  {journal} {\bibinfo  {journal} {Astrophys. J.}\ }\textbf {\bibinfo {volume} {178}},\ \bibinfo {pages} {347} (\bibinfo {year} {1972})}\BibitemShut {NoStop}%
\bibitem [{\citenamefont {{Chandrasekhar}}(1983)}]{chandra}%
  \BibitemOpen
  \bibfield  {author} {\bibinfo {author} {\bibfnamefont {S.}~\bibnamefont {{Chandrasekhar}}},\ }\href@noop {} {\emph {\bibinfo {title} {{The mathematical theory of black holes}}}}\ (\bibinfo {year} {1983})\BibitemShut {NoStop}%
\bibitem [{\citenamefont {Boyle}\ \emph {et~al.}(2019)\citenamefont {Boyle}, \citenamefont {Hemberger}, \citenamefont {Iozzo}, \citenamefont {Lovelace}, \citenamefont {Ossokine}, \citenamefont {Pfeiffer}, \citenamefont {Scheel}, \citenamefont {Stein}, \citenamefont {Woodford}, \citenamefont {Zimmerman} \emph {et~al.}}]{sxs_cat1}%
  \BibitemOpen
  \bibfield  {author} {\bibinfo {author} {\bibfnamefont {M.}~\bibnamefont {Boyle}}, \bibinfo {author} {\bibfnamefont {D.}~\bibnamefont {Hemberger}}, \bibinfo {author} {\bibfnamefont {D.~A.}\ \bibnamefont {Iozzo}}, \bibinfo {author} {\bibfnamefont {G.}~\bibnamefont {Lovelace}}, \bibinfo {author} {\bibfnamefont {S.}~\bibnamefont {Ossokine}}, \bibinfo {author} {\bibfnamefont {H.~P.}\ \bibnamefont {Pfeiffer}}, \bibinfo {author} {\bibfnamefont {M.~A.}\ \bibnamefont {Scheel}}, \bibinfo {author} {\bibfnamefont {L.~C.}\ \bibnamefont {Stein}}, \bibinfo {author} {\bibfnamefont {C.~J.}\ \bibnamefont {Woodford}}, \bibinfo {author} {\bibfnamefont {A.~B.}\ \bibnamefont {Zimmerman}}, \emph {et~al.},\ }\href@noop {} {\bibfield  {journal} {\bibinfo  {journal} {Classical and Quantum Gravity}\ }\textbf {\bibinfo {volume} {36}},\ \bibinfo {pages} {195006} (\bibinfo {year} {2019})}\BibitemShut {NoStop}%
\bibitem [{\citenamefont {Mroue}\ \emph {et~al.}(2013)\citenamefont {Mroue}, \citenamefont {Scheel}, \citenamefont {Szilagyi}, \citenamefont {Pfeiffer}, \citenamefont {Boyle}, \citenamefont {Hemberger}, \citenamefont {Kidder}, \citenamefont {Lovelace}, \citenamefont {Ossokine}, \citenamefont {Taylor} \emph {et~al.}}]{sxs_cat2}%
  \BibitemOpen
  \bibfield  {author} {\bibinfo {author} {\bibfnamefont {A.~H.}\ \bibnamefont {Mroue}}, \bibinfo {author} {\bibfnamefont {M.~A.}\ \bibnamefont {Scheel}}, \bibinfo {author} {\bibfnamefont {B.}~\bibnamefont {Szilagyi}}, \bibinfo {author} {\bibfnamefont {H.~P.}\ \bibnamefont {Pfeiffer}}, \bibinfo {author} {\bibfnamefont {M.}~\bibnamefont {Boyle}}, \bibinfo {author} {\bibfnamefont {D.~A.}\ \bibnamefont {Hemberger}}, \bibinfo {author} {\bibfnamefont {L.~E.}\ \bibnamefont {Kidder}}, \bibinfo {author} {\bibfnamefont {G.}~\bibnamefont {Lovelace}}, \bibinfo {author} {\bibfnamefont {S.}~\bibnamefont {Ossokine}}, \bibinfo {author} {\bibfnamefont {N.~W.}\ \bibnamefont {Taylor}}, \emph {et~al.},\ }\href@noop {} {\bibfield  {journal} {\bibinfo  {journal} {Physical Review Letters}\ }\textbf {\bibinfo {volume} {111}},\ \bibinfo {pages} {241104} (\bibinfo {year} {2013})}\BibitemShut {NoStop}%
\bibitem [{\citenamefont {Scheel}\ \emph {et~al.}(2025)\citenamefont {Scheel}, \citenamefont {Boyle}, \citenamefont {Mitman}, \citenamefont {Deppe}, \citenamefont {Stein}, \citenamefont {Armaza}, \citenamefont {Bonilla}, \citenamefont {Buchman}, \citenamefont {Ceja}, \citenamefont {Chaudhary} \emph {et~al.}}]{sxs_cat3}%
  \BibitemOpen
  \bibfield  {author} {\bibinfo {author} {\bibfnamefont {M.~A.}\ \bibnamefont {Scheel}}, \bibinfo {author} {\bibfnamefont {M.}~\bibnamefont {Boyle}}, \bibinfo {author} {\bibfnamefont {K.}~\bibnamefont {Mitman}}, \bibinfo {author} {\bibfnamefont {N.}~\bibnamefont {Deppe}}, \bibinfo {author} {\bibfnamefont {L.~C.}\ \bibnamefont {Stein}}, \bibinfo {author} {\bibfnamefont {C.}~\bibnamefont {Armaza}}, \bibinfo {author} {\bibfnamefont {M.~S.}\ \bibnamefont {Bonilla}}, \bibinfo {author} {\bibfnamefont {L.~T.}\ \bibnamefont {Buchman}}, \bibinfo {author} {\bibfnamefont {A.}~\bibnamefont {Ceja}}, \bibinfo {author} {\bibfnamefont {H.}~\bibnamefont {Chaudhary}}, \emph {et~al.},\ }\href@noop {} {\bibfield  {journal} {\bibinfo  {journal} {arXiv preprint arXiv:2505.13378}\ } (\bibinfo {year} {2025})}\BibitemShut {NoStop}%
\bibitem [{\citenamefont {Kelly}\ and\ \citenamefont {Baker}(2013)}]{Kelly12}%
  \BibitemOpen
  \bibfield  {author} {\bibinfo {author} {\bibfnamefont {B.~J.}\ \bibnamefont {Kelly}}\ and\ \bibinfo {author} {\bibfnamefont {J.~G.}\ \bibnamefont {Baker}},\ }\href {https://doi.org/10.1103/PhysRevD.87.084004} {\bibfield  {journal} {\bibinfo  {journal} {Phys. Rev. D}\ }\textbf {\bibinfo {volume} {87}},\ \bibinfo {pages} {084004} (\bibinfo {year} {2013})}\BibitemShut {NoStop}%
\end{thebibliography}%

\end{document}